\pdfoutput=1
\documentclass[prd,letter,nofootinbib
,twocolumn,preprintnumbers
]{revtex4}
\usepackage{slashed}
\usepackage{subfigure}
\usepackage{feynmp}
\usepackage{graphicx}
\usepackage{amsfonts}
\usepackage{color}

\def\hbar{\hspace{0pt}\raisebox{1pt}{$-$} \hspace{-7pt} h}

\def\5{\overline 5}

\newcommand{\be}{\begin{equation}}
\newcommand{\ee}{\end{equation}}
\newcommand{\bea}{\begin{eqnarray}}
\newcommand{\eea}{\end{eqnarray}}

\newcommand{\ba}{\begin{eqnarray}}
\newcommand{\ea}{\end{eqnarray}}
\newcommand{\no}{\nonumber}
\begin{document}
\title[]{Unitarity and Monojet Bounds on Models for DAMA, CoGeNT, and CRESST-II 
}

\author{Ian M. Shoemaker}
\email{ianshoe@lanl.gov}
\author{Luca Vecchi}
\email{vecchi@lanl.gov}
\affiliation{Theoretical Division T-2, MS B285, Los
Alamos
  National Laboratory, Los Alamos, NM 87545, USA}

\begin{abstract}
If dark matter interacts with quarks or gluons, the mediator of these interactions is either directly accessible at the LHC or is so heavy that its effects are encoded in contact operators. We find that the self-consistency of a contact operator description at the LHC implies bounds on the mediator scale stronger than those found from missing energy searches.  This translates into spin-independent elastic scattering cross-sections at a level $\lesssim 10^{-41}$ cm$^2$, with direct implications for the DAMA, CoGeNT, and CRESST-II anomalies.  We then carefully explore the potential of monojet searches in the light mediator limit, focusing on a $Z'$ model with arbitrary couplings to quarks and dark matter. We find that the Tevatron data currently provides the most stringent bounds for dark matter and $Z'$ masses below $100$ GeV, and that these searches can constrain models for the DAMA, CoGeNT, and CRESST-II anomalies only if the mediator can decay to a pair of dark matter particles.

\end{abstract}

\maketitle

\preprint{LA-UR-11-12298}

\section{Introduction}

If dark matter (DM) couples to quarks or gluons, its interactions can be discovered or constrained by LHC and Tevatron data in missing transverse energy (MET) searches. This has an interesting connection to direct detection experiments since the same couplings are probed. There is therefore the possibility that a one-to-one correspondence can be established between MET rates at colliders and scattering rates at direct detection experiments. Indeed, the status of DM direct detection, with the three experiments DAMA~\cite{Bernabei:2008yi}, CoGeNT~\cite{Aalseth:2011wp}, and CRESST-II~\cite{Angloher:2011uu} reporting excesses consistent with a 10 GeV DM candidate, cries out for an independent, astrophysics-free verification.  This has recently stimulated significant work aiming at a consistent explanation of these experiments~\cite{Frandsen:2011ts,Hooper:2011hd,Foot:2011pi,Belli:2011kw,Farina:2011pw,Schwetz:2011xm,Fox:2011px,McCabe:2011sr,An:2011ck,Fornengo:2011sz,Natarajan:2011gz,arXiv:1110.2721,Kelso:2011gd,Arina:2011zh}.

A rather general way of addressing this problem is the construction of an ``effective theory" of DM that can be probed at colliders~\cite{Cao:2009uw, Beltran:2010ww,DM1,Fermilab,DM2,DM3,Coolest, Mambrini:2011pw,Lin:2011gj,Kumar:2011dr}. If the mediators of quark-DM or gluon-DM interactions are sufficiently heavy, then their effects can be safely integrated out and described by higher-dimensional contact operators. In direct detection experiments the elastic scattering of WIMPs on nuclei is well-described by a contact interaction so long as the mediator mass is above a few MeV. In contrast, colliders like the LHC probe much higher momentum scales, and the validity of the effective description comes into question when the new physics is lighter than a few TeV. 

Here we will see that the very assumption that the DM interaction be consistently described by a contact operator up to LHC energies implies perturbative unitarity bounds on the mediator scale that turn out to be much stronger than those imposed by MET constraints. Importantly, this assumption is also shown to be at odds with both the DAMA and the uncontaminated (surface event-free) CoGeNT regions.

The implication of this result is that if DAMA, CoGeNT, and CRESST-II are indeed detecting DM, then the physics responsible for it must be directly accessible at the LHC and thus not generally amenable to a contact description. This gives hope to the prospect of confirming or ruling out the direct detection anomalies in the near term. To this end we analyze in detail constraints arising from monojet searches at the Tevatron and the LHC in models where the mediator is a new massive vector boson $Z'$ with arbitrary couplings to the standard model quarks and the DM. 

\section{Unitarity bounds on contact interactions}

To quantitatively assess the range of validity of a contact interaction at the LHC, consider adding to the SM Lagrangian the operator
\ba\label{contact}
\mathcal{O} = \frac{\overline{q}\gamma^\mu q~\overline{X}\gamma^\mu X}{\Lambda^2},
\ea
where $q$ is a SM quark and $X$ the dark matter, assumed to be a Dirac fermion for definiteness. Clearly, this is a very simplified picture: a complete, unambiguous description of the theory necessarily contains additional operators. These are expected to become relevant before the energy probed $E$ is such that the coupling $E^2/\Lambda^2$ becomes large. 
By na\"ive dimensional analysis, our simplified effective description obtained by considering \emph{only} the SM plus ${\cal O}$ is valid when $\Lambda> E/4\pi$.  

Although this simple consideration already provides a useful determination of the range of validity of our contact interaction assumption, it relies on a somewhat arbitrary definition of ``non-perturbativity."  

We can be slightly more precise regarding the validity of the $E/\Lambda$ expansion by requiring that our effective operator description be consistent with perturbative unitarity. (Analogous arguments are familiar from pion scattering, and have similarly been used to find, {\it e.g.}, an upper bound on the Higgs mass~\cite{FERMILAB-PUB-77-030-THY}, the scale of fermion mass generation~\cite{LBL-23951}, and on the dark matter mass~\cite{CFPA-TH-89-013}.) In the presence of $\mathcal{O}$, DM can be produced at the LHC via $pp\to X\overline{X}+\cdots$, where the dots stand for SM particles. The hard process contributing to these events is $q\overline{q}\to X\overline{X}$, and will be probed up to an energy $E \le 7$ TeV. The amplitude for the process $q\overline{q}\to X\overline{X}$ increases as a power of the center of mass energy $\sqrt{s}$ of the quark-antiquark system, and its probability eventually exceeds unity, signaling a loss of confidence in the contact interaction assumption. The strongest energy dependence arises in the scattering of opposite helicity, color-singlet states.
Specifically, the properly normalized initial and final states we consider are defined respectively as $\sum_\alpha(|q^\alpha_L\overline{q^\alpha_R}\rangle+|q^\alpha_R\overline{q^\alpha_L}\rangle)/\sqrt{2N_c}$, with $\alpha=1,\dots, N_c$ the color index, and $(|X_L\overline{X_R}\rangle+|X_R\overline{X_L}\rangle)/\sqrt{2}$, for which the (parton-level) amplitude reads
\ba
{\cal M}=2\sqrt{N_c}\frac{s}{\Lambda^2}\beta(s),
\ea
with $\beta(s)=\sqrt{1-4m_X^2/s}$ the dark matter velocity. Up to a normalization, this amplitude coincides with the zeroth order partial wave 
\ba\label{partial}
a_0(s)&=&\frac{1}{32\pi}\int^{+1}_{-1}d\cos\theta\,\,{\cal M}\\\no
&=&\frac{\sqrt{3}}{8\pi}\frac{s}{\Lambda^2}\beta(s).
\ea
Unitarity of the $S$-matrix imposes the bound $|a_0(s)|\le1/2\beta(s)$ for elastic as well as \emph{inelastic} processes (as a reference see, for example,~\cite{Marciano:1989ns}). We thus see that $\mathcal{O}$ provides a sensible, effective description of DM-quark interactions only if 
\ba\label{contactbound}
\Lambda &\gtrsim& 0.4~E \beta \no \\
 &\approx& 2.6\,~{\rm TeV}  \left(\frac{E \beta}{7~{\rm TeV}}\right),
\ea
where $\beta=\beta(E^2)$ and $E$ is the maximum energy flowing in the process.

\begin{figure}[t] 
\includegraphics[width=3.2in]{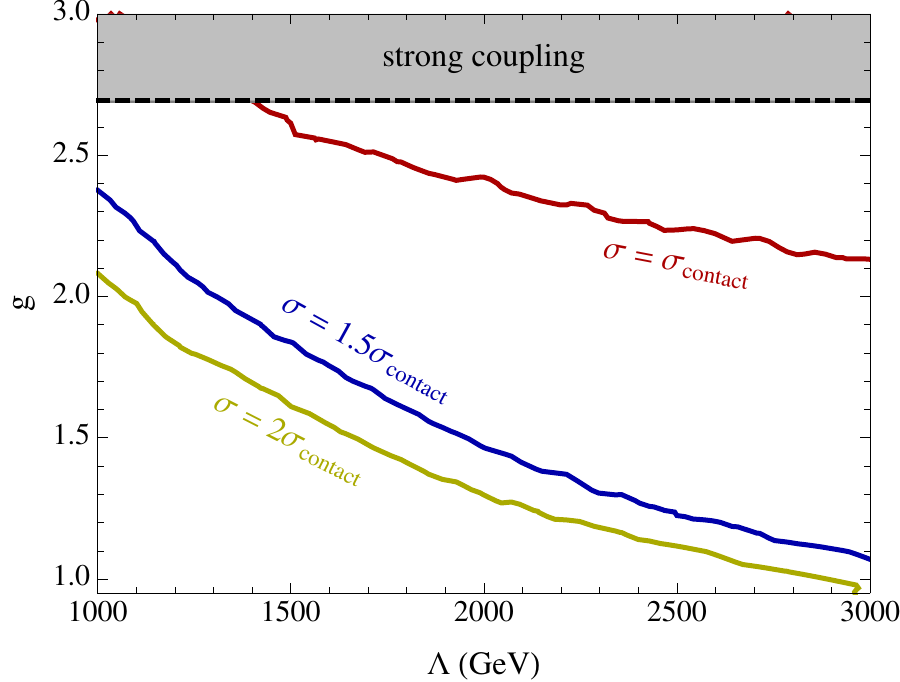}
\caption{\small Here we compare the monojet rate coming from the contact operator ${\cal O}$ ($\sigma_{contact}$) to that of a $Z'$ model ($\sigma$) with the $Z'$ coupling equally to quarks and dark matter. In both models we turn on only couplings to left-handed up quarks and dark matter and set $m_X=10$ GeV. In comparing the two models we fix $\Lambda = m_{Z'}/g$, such that their low-energy predictions agree. The shaded, strongly coupled region $ g^{2} > 4 \pi/\sqrt{3}$ represents the unitarity constraint of the UV model. \label{glambda}}
\end{figure}

\begin{figure}[t] 
\includegraphics[width=3.2in]{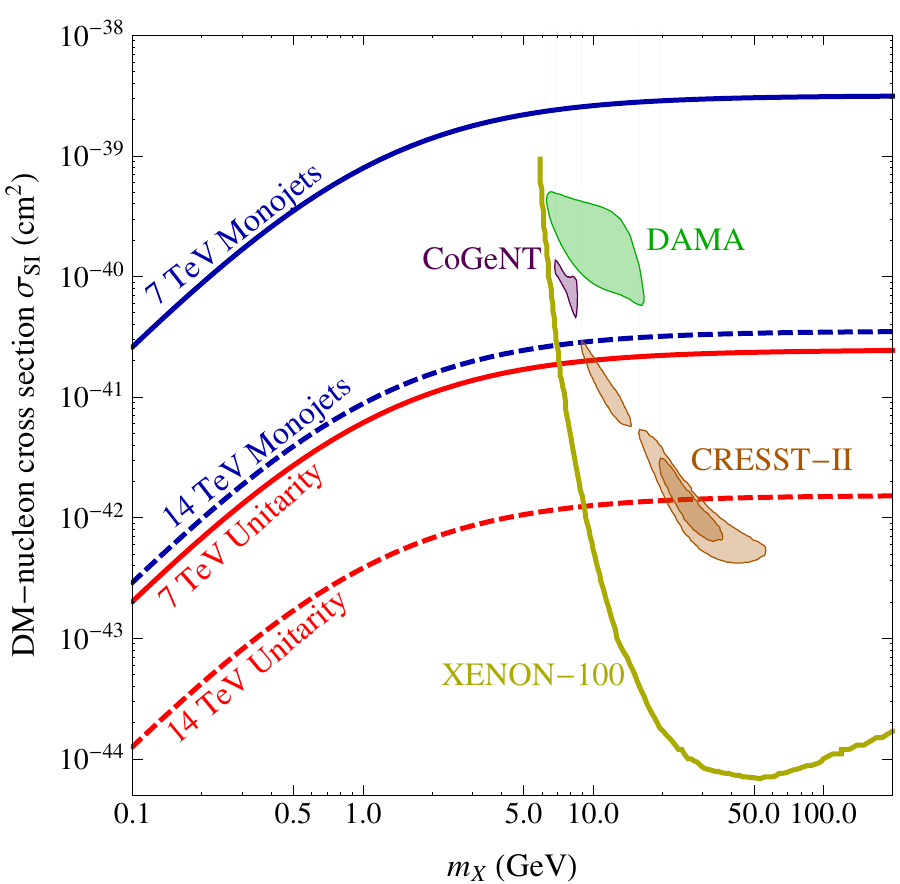}
\caption{\small Here we compare the direct detection upper bounds obtained from ATLAS monojets~\cite{DM2,DM3,Coolest} (solid blue) and the unitarity constraint~(\ref{contactbound}) with $E =$ 7 TeV (solid red line). We have also included the optimistic $5 \sigma$ LHC reach at 14 TeV~\cite{DM2} (dashed blue) along with the unitarity bound, at $E =$ 14 TeV (dashed red).  All bounds are derived under the assumption that the quark-DM interaction ${\cal O}$ remains contact at LHC energies and with universal quark coupling. For reference we include the DAMA $3\sigma$~\cite{Bernabei:2008yi,Savage:2008er}, CoGeNT 90$\%$ CL~\cite{Aalseth:2011wp}, and CRESST-II 1$\sigma$ and $2\sigma$ preferred regions~\cite{Angloher:2011uu} as well as the 90$\%$ CL XENON-100 bound~\cite{Aprile:2011hi}. \label{plotDD}}
\end{figure}

The physical interpretation of the above unitarity bound is that a UV-completion for $\cal{O}$ must appear \emph{before} Eq.~(\ref{contactbound}) is violated. This can manifest itself as new dynamics, or as higher order terms in the perturbative series. In either case our simplified description based only on $\cal{O}$ becomes unreliable when Eq.~(\ref{contactbound}) is violated, and a new, consistent theory of DM-quark interactions must be adopted. 

In calculations of $pp$ scattering at the $7$ TeV LHC one implicitly assumes a consistent and controllable description of quark interactions up to $\sim$ 7 TeV.~\footnote{In most searches, the kinematic cuts are typically of the order of a few hundred GeV, such that $E\sim 7$ TeV is justified.}  We therefore conservatively take $E = 7$ TeV in what follows.

 One may at first see this as an overly strong assumption since parton distribution functions (PDFs) will inevitably suppress the characteristic scale probed by ${\cal O}$ compared to the maximum center of mass energy of the LHC. However, although PDFs suppress the \emph{probability} that highly energetic $q\overline{q}\to X\overline{X}$ events are produced, they do not \emph{prohibit} them. Thus the conclusion that new, UV-physics beyond $\cal{O}$ must appear if $\Lambda \lesssim 2.6$ TeV is not altered by the effect of PDFs. Instead, the relevance of PDF suppression becomes now a practical question: Is this new physics actually observable? How well can DM-quark processes be approximated at the 7 TeV LHC by a contact operator description that violates Eq.~(\ref{contactbound})?

These questions cannot be addressed unambiguously, since for example they inevitably depend on the specific UV-completion of the DM-quark interaction, the process under consideration, and the uncertainties inherent in the experiment. However, it is instructive to investigate under which conditions a tractable UV-completion for DM-quark interactions can efficiently be described by ${\cal O}$ in a given collider search. We focus on monojet plus MET events, and without loss of generality consider models where ${\cal O}$ is completed by the exchange of a resonance of mass $g\Lambda$. Now the unitarity bound is equivalent to a perturbative bound on the coupling, $g^{2} < 4 \pi/\sqrt{3}$, and the assumption $\Lambda<2.6$ TeV is the statement that the mediator is kinematically accessible at the LHC. 

Clearly, if the mediator is exchanged at loop-level the coupling will be much smaller than unity and the resonance so light that a contact description is certainly inadequate. Yet, we will now argue that even for tree-level completions the contact operator description is typically inappropriate when the unitarity bound is violated. 

There are in general two classes of tree-level completions. The first consists of ``$s$-channel'' completions with interactions with a mediator $\phi$ of the form $\overline{q}q \phi$ and $\overline{X}X\phi$, and will be studied in detail in Section~\ref{zprime}. The second class of models are ``$t$-channel'' completions in which the mediator $\tilde\phi$ is a color-triplet and has interactions of the form $q\tilde\phi X$. In both cases, as soon as Eq.~(\ref{contactbound}) is violated the mediator becomes accessible at the LHC, and MET processes will tend to be dominated by resonance production (either $\overline{q}q \rightarrow \phi g \rightarrow \overline{X}X g$ or $q g \rightarrow X \tilde \phi \rightarrow q \overline{X}X$ depending on the particular completion), and the signal will typically be larger than that obtained from a contact interaction description. The resonance enhancement can be suppressed by pushing the theory towards the strong coupling limit $g^2\sim4\pi/\sqrt{3}$, but in so doing the theory becomes increasingly unpredictive. 

A quantitative measure of the impact of the resonance enhancement is shown in Fig.~\ref{glambda}, where we compare the monojet cross section of a contact description and an $s$-channel UV completion with the same value of $\Lambda$.  From this we see that $\Lambda$ above a few TeV is needed for an $s$-channel completion to produce a monojet signal compatible with that of a contact interaction. Similar results can be found in~\cite{DM3,Coolest}.

We therefore conclude that when $\Lambda$ is smaller than a few TeV the observable predictions of a generic, tractable UV-completion and a contact operator will typically differ greatly. A model-independent and conservative estimate of the bound is $\Lambda\lesssim2.6$ TeV.~\footnote{In a previous version of the paper we attempted a different and more conservative estimate. We now believe that any such estimate would necessarily have limited applicability and we therefore decided to take a more model- and process-independent approach.}



The consistency constraint Eq.~(\ref{contactbound}) is much more stringent than the bounds obtained so far from missing energy searches~\cite{DM1,Fermilab,DM2,DM3,Coolest}.\footnote{Monojets currently provide a direct constraint on contact DM-quark interactions of order $\Lambda \gtrsim 500$ GeV~\cite{DM1,Fermilab,DM2,DM3,Coolest}. Consistently with Fig.~\ref{glambda}, such low values of $\Lambda$ cannot be accounted for by tractable UV-completions; see also Fig.7 of~\cite{DM3} and Fig.7 of~\cite{Coolest}.} This continues to be true at 14 TeV, where unitarity requires $ \Lambda \gtrsim 5.2$ TeV, which is more stringent than the optimistic monojet bound with 100 fb$^{-1}$ of data at the 14 TeV LHC~\cite{DM2,Coolest}.

While in the above we chose to focus on the operator ${\cal O}$, it is easy to see that one would obtain similar results for operators with different Lorentz structure, as well as for dark matter with different spin. Analogous constraints also apply to dark matter interactions with gluons.


The consistency bound found above has important implications for direct searches of DM.   In Fig.~\ref{plotDD} we find that both the DAMA and uncontaminated CoGeNT regions are inconsistent with a contact operator description at the 7 TeV LHC, though the inclusion of surface events for CoGeNT may alter this conclusion~\cite{Collar,arXiv:1110.2721}. The direct detection implications of unitarity are even stronger in the case of velocity-dependent elastic scattering. In that case the bounds on the DM-nucleon scattering cross section in Fig.~\ref{plotDD} scale down parametrically by powers of the DM halo velocity ($v^{2} \sim 10^{-5}$). 

Thus, if DM is indeed the source of the DAMA and CoGeNT anomalies then the mediator responsible for DM-quark interactions is already kinematically accessible at the LHC.

\section{Dark matter and light $Z$'s at hadron colliders}
\label{zprime}

Once the contact interaction hypothesis is abandoned, matters become inevitably more model-dependent. Yet, irrespective of the details of the model, missing energy signals remain a characteristic signature of dark matter production at colliders. 

Jets and/or photons plus MET signals have been used before to bound dark matter~\cite{{DM1,Fermilab,DM2,DM3}} and neutrino interactions~\cite{Coolest}. Most of these efforts assume that the dark matter interacts with the SM via contact interactions. A first qualitative look at the effect of relaxing this assumption was presented in~\cite{Fermilab} (see also~\cite{arXiv:1111.2359,Coolest}), while a detailed analysis of the Tevatron monojet bounds for a specific model can be found in~\cite{Graesser:2011vj}. 

It would be useful to have a systematic and comprehensive study of the \emph{light mediator limit} that can easily be applied to one's favorite model. Here we aim to make a step in this direction by analyzing in detail models in which the interaction between dark matter $X$ and quarks $q$ is mediated by a new vector boson $Z'$.

\begin{figure}
\begin{center}
\includegraphics[width=3.2in]{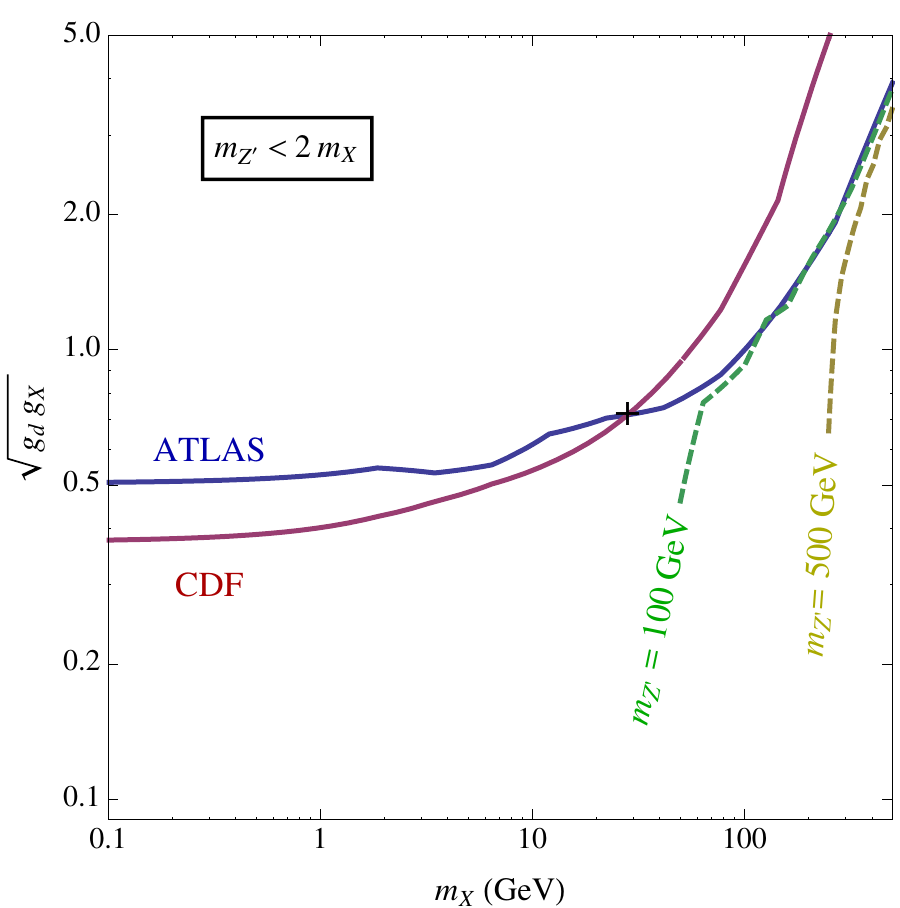}
\caption{\small Upper bounds on $\sqrt{g_{d} g_{X}}$ as a function of the DM mass when $m_{Z'} < 2 m_{X}$, with $m_{Z'} \ll 2m_{X}$ (solid lines), and $m_{Z'} =100$, 500 GeV (dashed lines). CDF provides the strongest bounds when the DM mass is less than about 30 GeV. Note that the value of the mediator mass only matters near threshold $m_{Z'} \sim 2 m_{X}$, where the effect of the $Z'$ width is also relevant. \label{plot0}}
\end{center}
\end{figure}

We find that for light $Z'$s and light DM the most constraining MET search currently arises from the \emph{monojet} data of CDF and the LHC:
\ba\label{j}
p\overline{p}/pp\to j+{\rm MET}.
\ea
In particular, we find the above monojet process to be more constraining than the analogous multijet + MET analyses. Moreover, the Tevatron's CDF data provide more stringent bounds than LHC data when the invariant mass of the DM system is small~\cite{Coolest}, as shown in Fig.~\ref{plot0}.  These results are simply a consequence of the fact that CDF's monojet selection criteria are softer and thus more sensitive to a MET signal coming from low mass physics.

\begin{figure*}[t] 
\begin{center}
\includegraphics[width=3.2in]{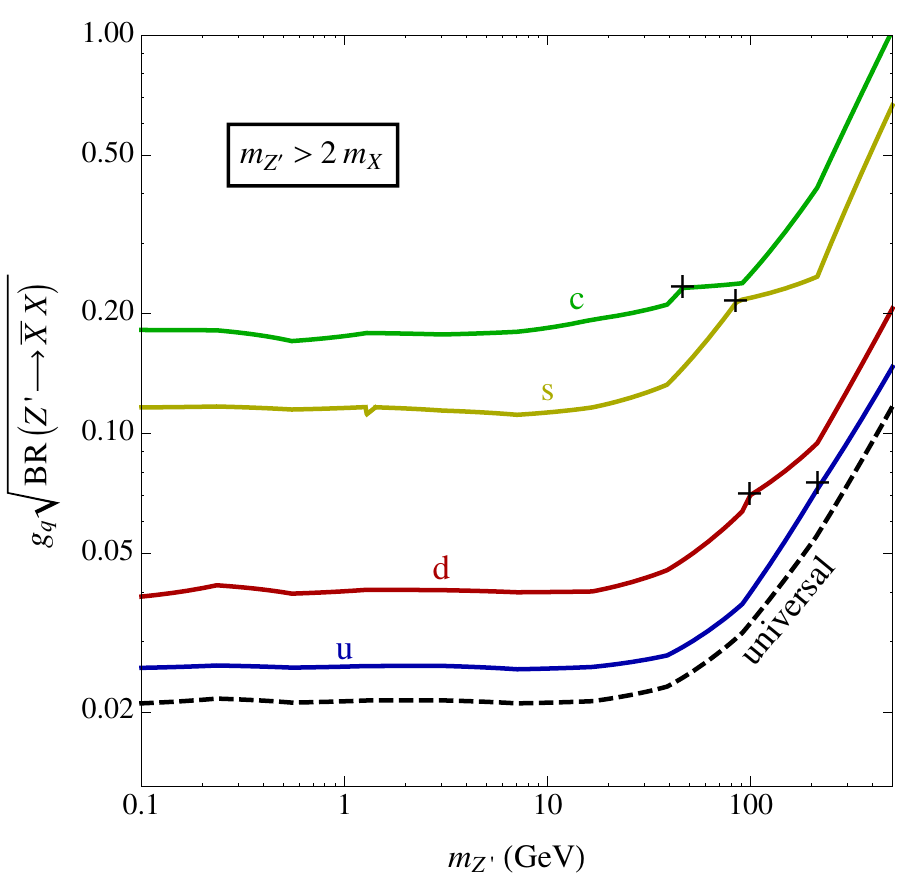}~~~~~\includegraphics[width=3.2in]{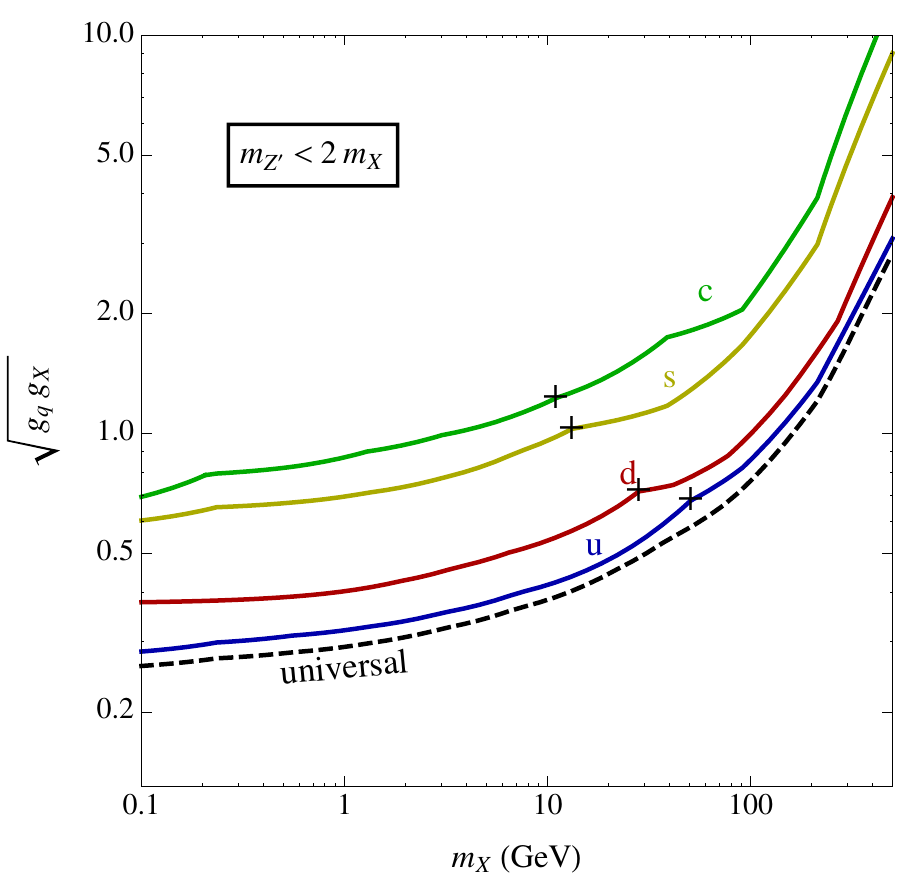}
\caption{\small $95\%$ CL monojet bounds on the $Z'$ model from CDF and ATLAS in the low DM mass and heavy DM regimes (see the text for more details). The bounds are derived assuming that a single quark coupling at a time is switched on. In Eq.~(\ref{gen}) we show how to modify the bounds in the most general case. As a reference we also quote the bound arising from the assumption of universal coupling. The ``plus" signs indicate the region were the CDF and ATLAS bounds cross each other (see also Fig.~\ref{plot0}).  \label{plot1}}
\end{center}
\end{figure*}

\subsection{Light $Z'$s and MET signatures}

There are at least two well motivated ways to couple a new light $Z'$ to the SM without affecting the huge phenomenological success of the SM~\cite{ArkaniHamed:2008qn,Batell:2009yf,Bjorken:2009mm,pospelov,Frandsen:2011cg,Graesser:2011vj,Cline:2011zr,Kang:2010mh,Mambrini:2010dq, Cline:2011uu}. The first is to couple the new boson via kinetic mixing with the hypercharge vector. The second is to promote the baryon number to a local symmetry. In either case the physics relevant here is effectively described by the SM Lagrangian plus ${\cal L}_{\rm kin}+Z'_\mu J^\mu$, where the first term contains pieces quadratic in $Z'$ and $X$, while the second describes the interactions between the two sectors. In general 
\ba
J^\mu=\sum_q\overline{q}\gamma^\mu(g^V_q+g_q^A\gamma^5)q+J^\mu_X,
\ea
where the precise structure of $J^\mu_X$ depends on the dark matter spin. Our MET bounds will be presented for fermionic dark matter, in which case 
\ba
J^\mu_X&=&\overline{X}\gamma^\mu(g^V_X+g_X^A\gamma^5)X.
\ea
We will discuss how the bounds change for dark matter of different spin at the end of this section.

(Note that even if the dark matter is assumed to be a SM singlet, loop effects can lead to couplings of $X$ to the SM vector bosons of the order $\sim10^{-2} g$. While these effects provide interesting model-dependent constraints on this scenario, they are irrelevant here.)

MET rates at hadron colliders are set by the following differential cross section
\ba
d\sigma_{\rm MET}=\sum_q d\sigma_q\,g_q^2g_X^2,
\ea
where the smallness of the SM quark masses allows us to define the effective quark coupling $g_q^2=|g^V_q|^2+|g^A_q|^2$. An analogous expression for the effective dark matter charge is also a good approximation when $m_X<500$ GeV. In this range we can thus write $g_X^2\approx |g^V_X|^2+|g^A_X|^2$.

For each quark the full parameter space is therefore described by four quantities: $m_{X},m_{Z'},\sqrt{g_{q}g_{X}},$ and the $Z'$ width $\Gamma_{Z'}$.  There are however two distinct physical regimes of interest where the number of parameters reduces: light dark matter, $2m_{X} < m_{Z'}$; and heavy dark matter, $2m_{X} > m_{Z'}$.  In the case of light dark matter, the mediator $Z'$ can be produced on-shell, and subsequently decay to a pair of DM particles.  Thus for a given $m_{Z'}$ MET searches provide an upper limit on the combination 
\be 
g_{q} \times \sqrt{{\rm BR}\left( Z' \rightarrow \overline{X}X \right)},
\ee
with a negligible dependence on the DM mass. (Here we have made use of the narrow-width approximation.) In contrast, for heavy dark matter the $Z'$ intermediate state is necessarily off-shell and for a given dark matter mass, MET searches then probe $\sqrt{g_{q} g_{X}}$, with a negligible dependence on $m_{Z'}$ and $\Gamma_{Z'}$.  The intermediate regime $2m_{X} \sim m_{Z'}$ is sensitive to all four parameters. This is illustrated in Fig.~\ref{plot0} where the $Z'$ mass dependence is apparent even in the heavy dark matter regime.  In order to present our bounds in simple, two-dimensional plots we ignore this intermediate, somewhat fine-tuned regime in what follows.

In the remainder we specialize to MET searches with a single jet in the final state, see~(\ref{j}), since we find these to be the most constraining.

\subsection{Monojet Bounds}

We simulate the signal using  {\texttt{Madgraph\_v5}}~\cite{MG5}, then apply the selection criteria of CDF~\cite{CDFADD2006,CDFADD2008,CDF2,Aaltonen:2012jb} and ATLAS~\cite{ATLAS1} using \texttt{Pythia 6.4}~\cite{Pythia} and cluster with \texttt{Fastjet 2.4.4}~\cite{Fastjet}.

In Fig.~\ref{plot1} we present the  95$\%$ CL bounds on DM-quark interactions in both the light (left) and heavy (right) DM regimes.  The limits are derived assuming that only one quark coupling at a time is switched on. When several couplings $g_q$ are turned on our bounds read
\ba\label{gen}
\sum_q\left(\frac{g_q}{g_q^{\rm bound}}\right)^n<1,
\ea
with $n=2(4)$ for the light (heavy) DM regime and $g_q^{\rm bound}$ being the bound on the quark flavor $q$ shown in Fig.~\ref{plot1}. The resulting bounds under the universal quark coupling assumption are shown as dashed black curves. Notice that the up and down quark bounds are stronger than the curves corresponding to the sea quarks as a result of PDF enhancement.  The curves asymptote to a constant when the DM invariant mass is small compared to the $p_{T}$ cut on the leading jet. 

At large DM invariant mass, the \texttt{veryHighPT} cut from ATLAS~\cite{ATLAS1} sets the strongest limit, whereas for low DM invariant mass the CDF 1 ${\rm fb}^{-1}$ analysis~\cite{CDF2} wins. We note that although a CDF 6.7 ${\rm fb}^{-1}$ analysis based on a jet $E_{T}$ analysis exists~\cite{Aaltonen:2012jb}, it is less constraining than the earlier 1$~{\rm fb}^{-1}$ counting analysis. 
  In Fig.~\ref{plot1} we have taken care to plot only the strongest of the collider monojet bounds.  The point at which the crossover occurs is shown as a ``plus'' in the figure.  Note that the mass at which the Tevatron starts setting the dominant bound is larger for quark flavors with a stronger PDF enhancement.   

Although we have focused on fermionic DM thus far, it is straightforward to translate the bounds to DM with other spins.  First, note that the left plot of Fig.~\ref{plot1} applies to {\it any} DM candidate so long as $Z' \rightarrow \overline{X} X$ is allowed.  For light DM, the bounds are roughly rescaled by the spin degree of freedom.  For example, for a complex scalar DM particle the bounds on $\sqrt{g_{X} g_{q}}$ in the right plot of Fig.~\ref{plot1} are decreased by a factor $1/\sqrt{2}$. (Here $g_X$ stands for the coupling of the $XXZ'$ cubic vertex.)

Lastly, we can translate our monojet bounds to direct detection cross section limits. As is visible from Fig.~\ref{plot1} the constraints on heavy DM are quite weak. For example, in the case of universally coupled $Z'$ with a $10$ GeV DM mass, the spin- and velocity-independent DM-nucleon elastic scattering cross section is bounded by
\ba
\sigma_{nX}\lesssim10^{-34}~{\rm cm}^2\left(\frac{20~{\rm GeV}}{m_{Z'}}\right)^4.
\ea
Note that in this regime the upper bound on the $Z'$ mass, $m_{Z'}<20$ GeV, implies that the constraint becomes {\it weaker} as the $Z'$ mass decreases.  In contrast, when $m_{Z'} > 2m_{X} = 20$ GeV the  limits are stronger, especially for narrow $Z'$s, implying 
\ba
\sigma_{nX}\lesssim10^{-36}~{\rm cm}^2\left(\frac{20~{\rm GeV}}{m_{Z'}}\right)^4\left(\frac{\Gamma_{Z'}/m_{Z'}}{10^{-2}}\right).
\ea
Now the bound for fixed $\Gamma_{Z'}/m_{Z'}$ becomes more stringent as the $Z'$ mass increases, approaching $\sigma_{nX}\lesssim10^{-40}$ cm${^2}$ when $m_{Z'}\sim500$ GeV.

\section{Conclusions}

We have seen that the very assumption that the DM-quark interaction remains contact at the LHC provides strong constraints on the scale of new physics. Specifically, perturbative unitarity arguments force the scale suppressing the higher dimensional operator to be above a few TeV. These bounds are notably stronger than those obtained from monojet searches at the LHC. 

In light of these results, the DAMA favored region and the (uncontaminated) CoGeNT 90$\%$ CL signal region \emph{cannot} be explained by interactions that remain contact at the 7 TeV LHC.  

We therefore relaxed the contact interaction assumption and analyzed in detail the constraints arising from monojet searches at the Tevatron and the LHC. 
We focused on models in which the DM-quark interaction is mediated by a light $Z'$ with generic couplings to the SM quarks and the DM, and presented the bounds in a way that our results are readily applicable to the reader's favorite scenario.

We find that the strongest bounds are currently provided by the Tevatron's CDF experiment for DM and $Z'$ masses below $\sim100$ GeV, with the LHC taking over for heavier masses.

Translating our monojet bounds to limits on the DM-nucleon elastic scattering cross section we find that these searches can constrain models for the DAMA, CoGeNT, and CRESST-II excesses only if $Z' \rightarrow \overline{X} X$ is allowed and if the $Z'$ is a narrow resonance.

\vspace{1.2cm}

\begin{center}
{{\bf Note Added}}
\end{center}

After the submission of our arXiv paper, the preprint \cite{Fox:2012ee} appeared with a discussion claiming that the unitarity constraint of the present paper is in practice unimportant for rather low cutoffs. This conclusion is based on the inspection of the $\overline{X}X$ invariant mass distribution for a MET search, in the case of a contact operator with $\Lambda \sim 600$ GeV. From this the authors find that the number of events that ``violate unitarity" (i.e., those events with $m_{\overline{X}X} > \Lambda/0.4 \approx 1.5$ TeV) is negligible for light dark matter. However, as we clarified above, the unitarity bound implies that a consistent description requires that new physics appear at invariant masses \emph{below} $1.5$ TeV. A more telling measure of the validity of the contact operator description would be a comparison between a contact operator and a UV-complete theory leading to the same $\Lambda$. This comparison was already made in~\cite{DM3} for monojets, where it was shown that in order for an $s$-channel completion to agree with the contact interaction prediction one should have couplings of order $g\sim5-10$, in agreement with our Fig.~\ref{glambda}. With such large couplings, yet higher dimensional operators are expected to become relevant and the monojet prediction is plagued by $O(1)$ uncertainties. We interpret this as a confirmation of our conclusions.

\acknowledgements

We would like to thank Alexander Friedland, Michael Graesser, and Hasan Y\"{u}ksel for illuminating conversations. We are especially grateful to Alex Friedland for providing us with the scripts used to scan the $Z'$ parameter space, to Michael Graesser for discussions on partial wave unitarity, and to Hasan Y\"{u}ksel for providing useful Mathematica expertise.  This work
was supported by the DOE Office of Science and the LANL LDRD program.


\end{document}